\documentclass{aastex}
\usepackage{emulateapj5}

\shorttitle{Atmospheres of TrES-1 and HD209458b}
\shortauthors{Fortney et al.}

\newcommand{\T}{TrES-1}
\newcommand{\hd}{HD209458b} 
\slugcomment{\bf}
\slugcomment{Accepted to ApJ Letters, May 17, 2005}

\begin{document}

\title{Comparative Planetary Atmospheres: Models of TrES-1 and HD209458b}

\author{J. J. Fortney\altaffilmark{1},  M. S. Marley\altaffilmark{1}, K. Lodders\altaffilmark{2}, D. Saumon\altaffilmark{3}, R. Freedman\altaffilmark{1}$^,$\altaffilmark{4}} 

\altaffiltext{1}{NASA Ames Research Center, Space Science and Astrobiology Division, MS 245-3, Moffett Field, CA 94035;
jfortney@arc.nasa.gov, Mark.S.Marley@nasa.gov, freedman@darkstar.arc.nasa.gov}
\altaffiltext{2}{Planetary Chemistry Laboratory, Department of Earth and Planetary Sciences, Washington University, St. Louis, MO 63130;
lodders@levee.wustl.edu}
\altaffiltext{3}{Los Alamos National Laboratory, MS F699, Los Alamos, NM 87545;
dsaumon@lanl.gov}
\altaffiltext{4}{SETI Institute, 515 N. Whisman Road, Mountain View, CA 94043}

\begin{abstract}

We present new self-consistent atmosphere models for transiting planets \T~and \hd.  The planets were recently observed with the \emph{Spitzer Space Telescope} in bands centered on 4.5 and 8.0 $\mu$m, for \T, and 24 $\mu$m, for \hd.  We find that standard solar metallicity models fit the observations for \hd.  For \T, which has an $T_{\mathrm{eff}}\sim300$ K cooler, we find that models with a metallicity 3-5 times enhanced over solar abundances can match the 1$\sigma$ error bar at 4.5 $\mu$m and 2$\sigma$ at 8.0$\mu$m.  Models with solar abundances that included energy deposition into the stratosphere give fluxes that fall within the 2$\sigma$ error bars in both bands.  The best-fit models for both planets assume that reradiation of absorbed stellar flux occurs over the entire planet.  For all models of both planets we predict planet/star flux ratios in other \emph{Spitzer} bandpasses.

\end{abstract}



\section{Introduction}
The detection of photons emitted by an extrasolar planet is a landmark feat in the history of astronomy.  Ten years after the initial discovery of the first extrasolar giant planet (EGP) 51 Peg b \citep{Mayor95}, \citet{Charb05} report a dectection of infrared flux from \T~and \cite{Deming05b} report a similar detection for \hd.  Both observations were made with the \emph{Spitzer Space Telescope} as each planet passed behind its parent star.  While a number of papers have aimed at predicting the optical and infrared spectra of hot EGPs \citep{SS98, Marley99, Barman01, Sudar03}, these results are the first infrared detections (rather than upper limits) that can be compared with models. 

\T, discovered by \citet{Alonso04}, is similar to the well-known planet \hd~\citep{Charb00,henry00} in mass and orbital period but it receives less stellar irradiation and has a significantly smaller radius.  \citet{Sozzetti04} and \citet{Laughlin05} have recently placed tighter constraints on the planet's mass and radius by improving our knowledge of the physical parameters of the parent star.

\T~and \hd~are members of the class of ``Pegasi planets'' (or ``hot Jupiters'') that orbit their parent stars at less than $\sim$~0.1 AU and are presumed to be tidally locked.  Unlike brown dwarfs of similar effective temperature, the luminosity of Pegasi planets is dominated by absorbed and re-radiated stellar flux, rather than intrinsic luminosity due to cooling of its interior.  Atmospheric dynamics will redistribute this absorbed energy around the planet with an efficiency that is currently unknown \citep{Showman02, Cho03, Burkert05, Cooper05}.  Because of the dominance of absorbed flux over intrinsic flux, the planet's effective temperature $T_{\mathrm{eff}}$, is equal to its equilibrium temperature, $T_{\mathrm{eq}}$, given by:
\begin{equation}
\label{Teq}
T_{eq}^4 = f(1-A) L_{\star}/(16 \pi \sigma d^2),
\end{equation}
where $f$ is 1 if the absorbed radiation is able to be radiated away over the entire planet ($4\pi$ steradians) or 2 if it only radiates on the day side ($2\pi$ steradians).  $A$ is the planet's Bond albedo, $L_{\star}$ is the luminosity of the star, $\sigma$ is the Stefan-Boltzmann constant, and $d$ is the planet's orbital distance.

\section{Review of the Observations}
\citet{Charb05} used the Infrared Array Camera (IRAC) aboard \emph{Spitzer} to observe the emission of the combined \T~planet/star system. The timing of the observations included time before, during, and after the secondary eclipse (when the planet's light is blocked by the star).  The observations were performed simultaneously in bands 2 and 4, centered on 4.5 and 8.0 $\mu$m, respectively.  The planet+star system was indeed brighter in both bands when the planet was visible, indicating thermal emission from the planet was detected.  The observed eclipse depths, in units of relative flux, were 0.00066 $\pm$ 0.00013 at 4.5 $\mu$m and 0.00225 $\pm$ 0.00036 at 8.0 $\mu$m.  The \hd~observations by \citet{Deming05b} utilized this same method, except that their observation used the Multiband Imaging Photometer for Spitzer (MIPS) instrument at 24 $\mu$m.  The observed flux ratio was 0.00260 $\pm$ 0.00046.

From their observations \citet{Charb05} derived brightness temperatures in each of the two bands, $T_{4.5}=1010\pm 60$ K and $T_{8.0}=1230\pm 110$ K.  It is important to notice that $T_{8.0}>T_{4.5}$, which was not predicted by models of Pegasi planets cited above, nor for brown dwarfs of similar $T_{\mathrm{eff}}$ \citep{Marley96,Burrows97,Allard01,SML03}, nor is it observed in the mid infrared spectra of brown dwarfs to date \citep{Roellig04,Patten04}.  \citet{Charb05} also derive a planetary $T_{\mathrm{eff}}=1060 \pm 50$ K and $A=0.31 \pm 0.14$, but as the planet is not a blackbody, the values found are only suggestive.  For \hd, \citet{Deming05b} determine a brightness temperature $T_{24}=1130\pm 150$ K.

\section{Methods}
To obtain our atmospheric pressure-temperature (\emph{P-T}) profiles and spectra for the planets we employ a 1D  model atmosphere code that has been used for a variety of planetary and substellar objects.  The code was first used to generate profiles and spectra for Titan by \citet{Mckay89}.  It was significantly revised to model the atmospheres of brown dwarfs \citep{Marley96, Burrows97, Marley02}, Uranus \citep{MM99}, and EGPs \citep{Marley98}.  It explicitly includes both incident radiation from the parent star and thermal radiation from the planet's atmosphere.  The basic radiative transfer solving scheme was developed by \citet{Toon89}.  We use the elemental abundance data of \citet{Lodders03} and compute chemical equilibrium compositions following \citet{Fegley94} and \citet{Lodders02}.  In addition we maintain a large and constantly updated opacity database.  We predict all cloud properties using the model of \citet{AM01} with a sedimentation efficiency parameter $f_{sed}=3$,  which fits spectral observations of cloudy L-dwarfs \citep{Marley02}.  This model places 90\% of the optical depth of a cloud within 1 scale height of the cloud base.  Further details can be found in \citet{Marley02} and Marley et al.~(in prep).

We model the impinging stellar flux from 0.26 to 10.0 $\mu$m and the emitted thermal flux from 0.26 to 325 $\mu$m.  All the relevant planetary parameters for \T~are taken from \citet{Sozzetti04}.  Those for \hd~are taken from \citet{Brownetal01}.  For the \T~stellar model we use the Kurucz K0V model atmosphere computed for the \citet{Charb05} paper.  For \hd~we use a \citet{Kurucz93} G0V model (L$_{\star}=1.6$L$_{\odot}$) with parameters described in \citet{Brownetal01}.  The planet's radiative-convective \emph{P-T} profile is arrived at iteratively until the net flux is conserved to at least one part in $10^6$.  We compute all band-averaged flux density ratios using Eq.~(1) from \citet{Charb05}.
%
%
Our model atmosphere code computes the \emph{P-T} profile and low resolution spectra covering our full wavelength range.  To generate a high-resolution spectrum we take the generated \emph{P-T} profile and use a full line-by-line radiative transfer code, using the same chemistry and opacity database used in determining the \emph{P-T} profile and low-resolution spectrum \citep{Saumon00}.

\section{Results}
\subsection{Atmospheric Pressure-Temperature Profiles}
\begin{figure}
\plotone{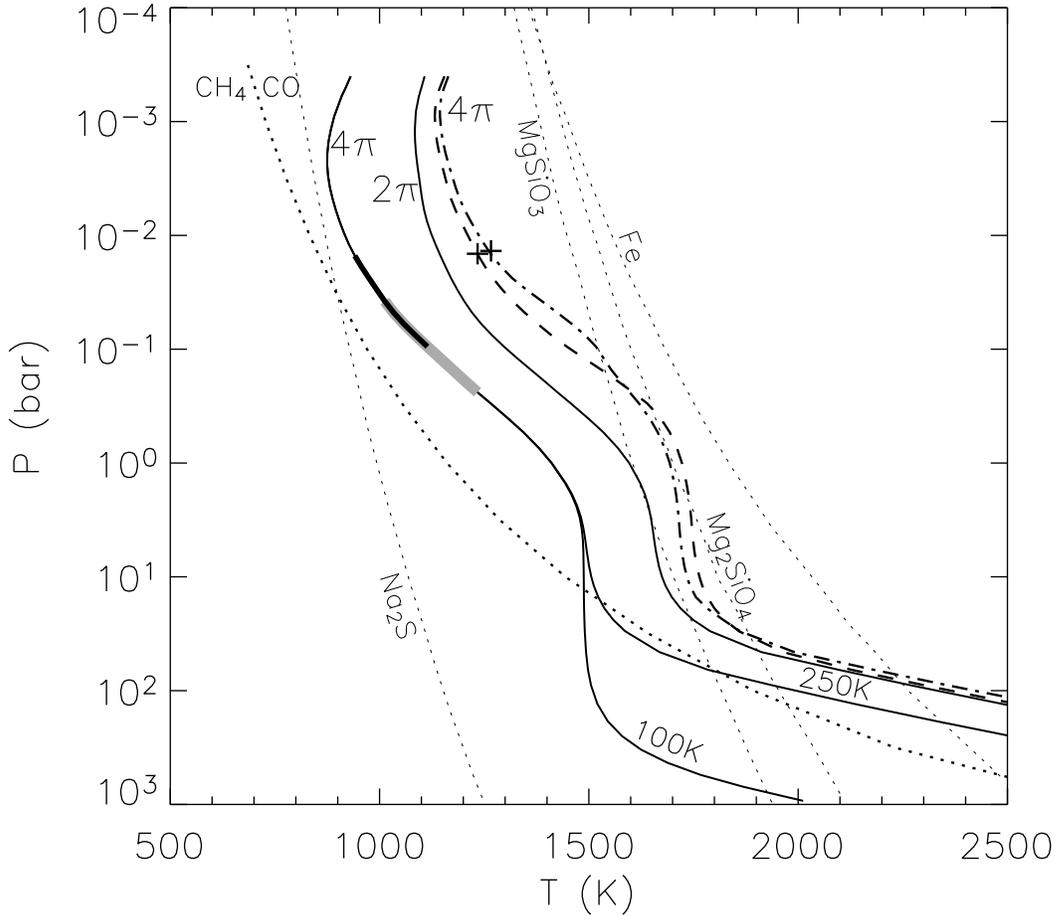}
\caption{Pressure-temperature (\emph{P-T}) profiles for \T~and \hd.  The solid lines are two profiles for \T~assuming either $2\pi$ (hotter) or $4\pi$ (cooler) reradiation of absorbed stellar flux.  Both profiles assume $T_{\mathrm{int}}=250$ K.  For the $4\pi$ case, we also plot a profile with  $T_{\mathrm{int}}=100$ K.  For the $4\pi$ profile the thick black portion of the profile shows the extent of the brightness temperatures in the 8.0 $\mu$m band.  The even thicker gray portion of the profile shows the brightness temperatures in the 4.5 $\mu$m band.  We plot two $4\pi$ profiles for \hd.  The dashed line is cloud-free and the dash-dot line includes the opacity of MgSiO$_3$ and Fe clouds.  The model brightness temperatures at 24 $\mu$m are marked with plusses.  The thick dotted line is the boundary where CO and CH$_4$ have the same abundance.  Condensation curves for Na$_2$S, MgSiO$_3$, Mg$_2$SiO$_4$, and Fe are thin dotted lines.  
\label{figure:pt}}
\end{figure}

Our computed atmosphere profiles are shown in \mbox{Figure~\ref{figure:pt}}.  Two cloudless profiles for \T~are shown, one under the assumption that absorbed stellar radiation is reradiated from the entire planet, and one for the assumption that the planet can only reradiate this energy on the day side.  For these cases we derive a $T_{\mathrm{eff}}$ of 1134 K and 1352 K, respectively.  Both of these profiles assume solar metallicity and an intrinsic temperature, $T_{\mathrm{int}}$ of 250 K.  This $T_{\mathrm{int}}$ is the $T_{\mathrm{eff}}$ the planet would have if it were in isolation.  For the $4\pi$ case we also plot a profile for $T_{\mathrm{int}}=100$ K, which leads to a deeper radiative zone, but has no effect on the emitted spectrum.  The presence of an outer radiative zone is predicted for the atmospheres of all Pegasi planets and is a caused by the dominance of external (stellar) radiation over internal heat flux \citep{Guillot96}.  The planet's current $T_{int}$ can only be derived from thermal evolution models of the planet, which we do not compute here.  For the $4\pi$ \T~profile we show the model's range of brightness temperatures in 4.5 and 8.0 $\mu$m bands.

We also plot our derived \emph{P-T} profile for \hd.  This profile assumes $4\pi$ reradiation, solar metallicity, and $T_{\mathrm{int}}=250$ K.  For \hd~we find $T_{\mathrm{eff}}=1442$ K, $\sim$~300 K hotter than that for \T.  The model's brightness temperature at 24 $\mu$m is shown as a plus on this profile.  We also plot the boundary where CO and CH$_4$ have the same abundance.  We predict CO will be the dominant (but not exclusive) carrier of carbon for both planets.  Also shown are the condensation curves for Na$_2$S, enstatite (MgSiO$_3$), forsterite (Mg$_2$SiO$_4$) and iron (Fe).  Clouds form deep in the atmosphere of \T~and therefore have no effect on the planet's spectrum.  For \hd, we find that a potential enstatite cloud layer does affect the \emph{P-T} profile and spectrum, as shown by the dash-dot profile.  High clouds in the atmosphere of \hd~are compatible with the available data from atmospheric transmission spectroscopy \citep{Charb02,Fortney03,Deming05a}, although the clouds in those studies are at mbar pressures.  Our profile for \hd~is very similar to the $4\pi$ model of \citet{Iro05}, although our profiles show hints of stratospheric temperature inversions due to absorption by near infrared water bands, which to our knowledge other authors have not found.

\subsection{Flux Ratios and Brightness Temperatures}
Our computed flux ratios for the \hd~model atmospheres match the \citet{Deming05b} observation well.  We find $\Delta F = 0.00305$ for a cloudless model or with clouds of iron and forsterite, which is within the 1$\sigma$ error bar.  If the silicate cloud is not forsterite, but enstatite, which forms higher in the atmosphere, we find $\Delta F = 0.00316$.  Models computed assuming 2$\pi$ reradiation would be brighter at 24 $\mu$m, leading to a poorer fit to the datum.

\begin{figure}
\epsscale{1.2}
\plotone{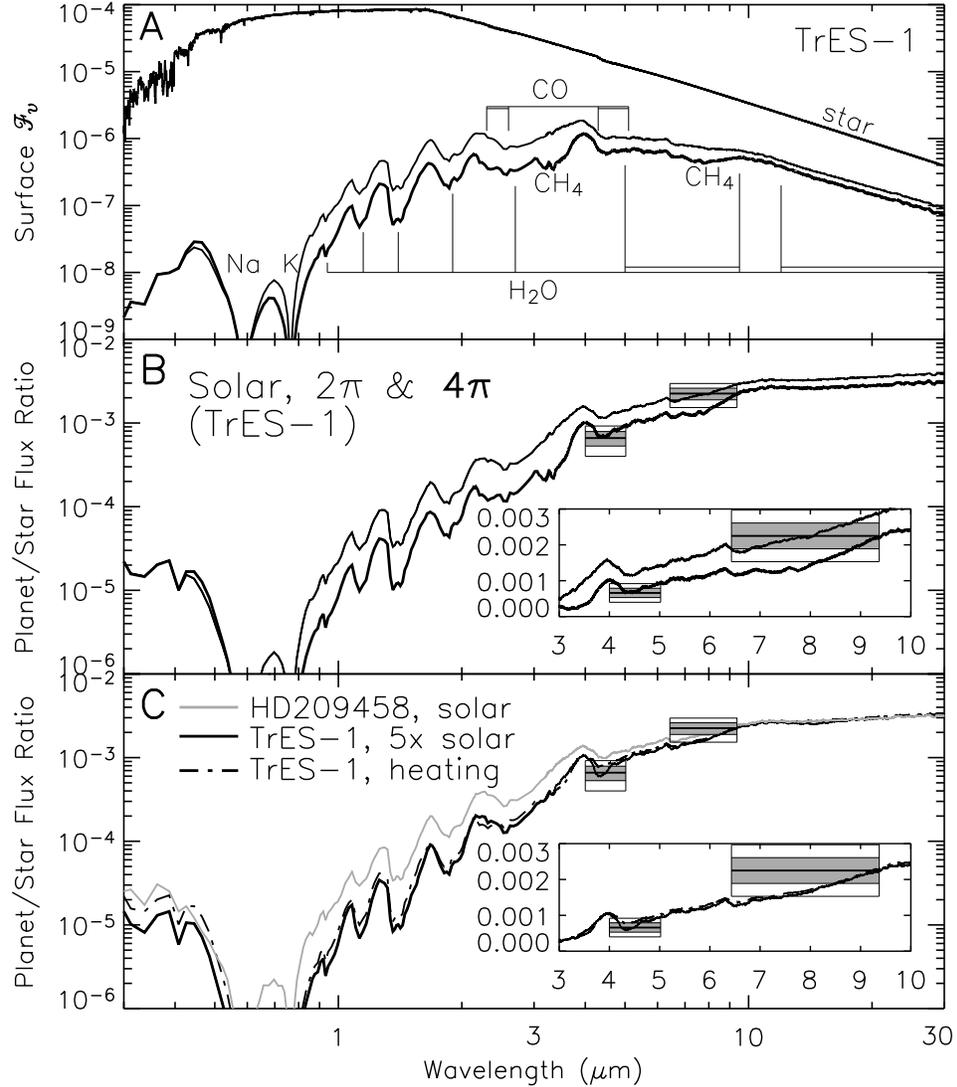}
\caption{Panel A: Emergent surface flux density (in erg s$^{-1}$ cm$^{-2}$ Hz$^{-1}$) for the TrES-1 K0V parent star and two planet models---the higher curve assumes $2\pi$ flux reradiation and the lower thick curve $4\pi$.  Relevant absorption features are labeled.  Panel B: Planet/star flux density ratios for the models in panel A, as viewed by a distant observer.  The full-width at half-maximum wavelength range of both bands are shown.  The one- and two-$\sigma$ error bars of the \citet{Charb05} measurements are shown as shaded boxes and empty boxes, respectively.  The inset is a blowup of the 3-10 $\mu$m region.  Panel C: The gray curve is HD209458, the solid black curve is \T~with 5 times solar metallicity, and the dash-dot curve is \T~with solar metallicity and a heated upper atmosphere. (See text.)
\label{figure:spec3}}
\end{figure}

For \T~we will examine our model atmospheres in detail. In \mbox{Figure~\ref{figure:spec3}} we show three panels of results.  Panel A shows the surface flux of the model K0V stellar spectrum along with two computed surface fluxes of planet \T.  These spectra are for $2\pi$ and $4\pi$ reradiation and solar metallicity.  As expected, the $2\pi$ spectrum is brighter.  Note that between 4-10 $\mu$m we find that emitted planetary flux is $\sim$~7-9 orders of magnitude brighter than reflected stellar flux.

The lower two panels of \mbox{Figure~\ref{figure:spec3}} show the planet/star flux ratio, as would be seen by a distant observer, which folds in the factor of $\sim$~60 difference in surface area of the planet and star.  These panels compare our computed ratios to those measured by \citet{Charb05}.  For $2\pi$ reradiation our ratio at 4.5 $\mu$m is 4.5$\sigma$ too high, indicating the planet model is far too luminous at these wavelengths.  However, the fit at 8.0 $\mu$m is quite good.  For the $4\pi$ model, the flux at 4.5 $\mu$m is a reasonable fit, falling just above 1$\sigma$ error bars, while at 8.0 $\mu$m the model falls below the 2$\sigma$ error bar.  Both of these models have approximately the same infrared spectral slope, which is obviously not as steep as \citet{Charb05} observed.  This slope is mainly controlled by  H$_2$O absorption on the blue side of the 4.5 $\mu$m band and strong CO absorption on the red side.  At 8.0 $\mu$m, H$_2$O and CH$_4$ are the dominant absorbers.  Based on these results, we rule out the 2$\pi$ model and examine the sensitivity of the spectral slope to model parameters.  Our computed brightness temperatures for the $4\pi$ models are $T_{4.5}=1075$ K and $T_{8.0}=995$ K, while \citet{Charb05} found $T_{8.0}>T_{4.5}$.  Compared to a blackbody, the planet is redder while the models are bluer.

A flux deficit at 4.5 $\mu$m, relative to atmosphere models using equilibrium chemistry, is a common feature of Jupiter and T dwarfs \citep{Golim04}. This is likely due to dredging of CO from deeper layers, which absorbs flux from 4.5-5 $\mu$m \citep{Fegley96,Saumon03}.  We find this process cannot affect the atmosphere of \T~as CO is already the dominant carbon-bearing molecule in the planet's atmosphere.

We also varied the metallicity of the planet's atmosphere.  We computed chemistry and opacity grids at metallicities of 3.2 ([M/H]=0.5) and 5.0 ([M/H]=0.7) times solar.  Recall that Jupiter's atmosphere is enhanced in heavy elements by factors of $\sim$~2-4.  With 3.2$\times$ or 5$\times$ solar metallicity the models fall within the 1$\sigma$ error bar at 4.5 $\mu$m and 2$\sigma$ at 8.0 $\mu$m.  Panel C of \mbox{Figure~\ref{figure:spec3}} shows the flux ratios for the 5$\times$ solar abundance model.  Increasing the metallicity by 3.2$\times$ decreases the CH$_4$/CO ratio by an order of magnitude. (The position of CH$_4$/CO=1 curve varies inversely proportional to the metal/hydrogen ratio.)  This in turn increases the 8.0 $\mu$m band flux by weakening the strong CH$_4$ band at 7.8 $\mu$m.  Increasing the metallicity also increases the amount of H$_2$O, which decreases the flux on the blue side of the 4.5 $\mu$m band.

We also considered whether we could further improve the fit by increasing the C/O ratio from 0.5 (solar) to 0.7, therefore increasing CO absorption at 4.5 $\mu$m and decreasing H$_2$O absorption at 8.0 $\mu$m.  However, increasing the C/O ratio (by increasing the carbon abundance) \emph{increases} the CH$_4$ abundance relative to CO (and H$_2$O) and flux actually decreases in the 8.0 $\mu$m band.  Increasing [M/H] from 0 to +0.5 or increasing the C/O ratio from 0.5 to 0.7 has very little effect on the strength of the CO band.  This is because the band is formed very high in the atmosphere and is already saturated at solar abundances.

Another avenue we pursued was increasing the temperature of the upper atmosphere.  In our models the pressures at which optical depth unity occurs in the 8.0 $\mu$m band are slightly smaller than those in the 4.5 $\mu$m band, although there is considerable overlap. (See \mbox{Figure~\ref{figure:pt}}.) Therefore, increasing the temperatures at lower pressures preferentially increases the $T_{8.0}$ relative to $T_{4.5}$.  We deposited additional energy in the form of a Chapman function \citep[e.g.][]{ChambHunt}, into the upper atmosphere at pressures ranging from 1 to 10 mbar.  Possible unmodeled energy sources include conduction downward from a hot thermosphere \citep{Yelle04}, the breaking of gravity waves, or an unmodeled optical absorber, such as photochemical products.  Profiles that warm the atmosphere at 10 mbar by $\sim$~200 K (requiring $\sim$~$10^7$ erg cm$^{-2}$ sec$^{-1}$, or 2.5\% of the incoming stellar flux) generate more flux in the 8.0$\mu$m band, and reach the 2$\sigma$ error bars in both bands.  Reprentative flux ratios of one of these models is shown in Panel C of \mbox{Figure~\ref{figure:spec3}}.

In another trial (not shown) we modeled a strong continuum opacity source from 2-6 $\mu$m.  An appropriately strong absorber cuts down the flux in the 4.5 $\mu$m band while increasing flux in the 8.0 $\mu$m band, which allows the model to fit the 1$\sigma$ error bars in both bands.  It is unclear what this mystery opacity source could be.  \citet{Liang04} found that \hd~should be free of hydrocarbon hazes as atmospheric temperatures are too high for hydrocarbons to condense.  Their conclusion should hold for \T, although it is $\sim$~300 K cooler.  We note that the photochemical destruction of CH$_4$ \citep{Liang03} would lead to slightly weaker absorption at 8.0 $\mu$m.

\section{Discussion}
We find that a solar metallicity atmosphere for \T~marginally fits the \emph{Spitzer} observations.  Models with enhanced metallicity are promising as they fall within the 1$\sigma$ error bar in the 4.5 $\mu$m band and 2$\sigma$ error bar in the 8.0 $\mu$m band.  Solar metallicity models with extra energy deposition into the stratosphere fit at the 2$\sigma$ level.  As \mbox{Table~\ref{table:param}} shows, observations with IRAC at 3.6 $\mu$m look most promising for further separating the predictions of standard solar abundance models from  others.  For the model phase space we explore it is extremely difficult to obtain brightness temperatures of $T_{8.0}>T_{4.5}$.  One possible remedy would be for the planet to have a cooler troposphere than we predict here, but a much hotter stratosphere, which could cause the planet's absorption features to become emission features, and vice versa.  We can recover this behavior with an ad-hoc \emph{P-T} profile.  We note that in Jupiter the 7.8 $\mu$m CH$_4$ feature is seen in emission, not absorption.  

As shown above, standard solar abundance models fit the \citet{Deming05b} MIPS observation well.  However, with this one data point \emph{metal enhanced atmospheres are not excluded}.  The known differences between \hd~and \T~are the former's larger radius and higher $T_{\mathrm{eff}}$.  It is not yet clear if these atmospheres are substantially different in character.  Possibilities include opacity due to silicate condensates in \hd~but not \T, and opacity due to photochemical products in \T~but not \hd.  \emph{Spitzer} observations in additional bandpasses for both planets will help to characterize these atmospheres.  Observations of the depth of the primary eclipse with \emph{Spitzer} will allow the measurements of nightside fluxes of the planet, which would place strong constraints on atmospheric dynamics and on the energy redistribution between the day/night hemispheres.  In \mbox{Table~\ref{table:param}} we collect our computed planet/star flux ratios for a variety of \T~and \hd~models in all four IRAC bands and the MIPS 24 $\mu$m band.

Since models of the mid infrared spectra of brown dwarfs match observations to date \citep{Roellig04,Patten04}, we suggest that while atmospheric composition plays a role, a complete explanation for the red spectral slope of \T~likely will involve energy from the parent star.  Whether this is due to unknown photochemical processes, unmodeled energy sources that dramatically modify the \emph{P-T} profile, or some other process is not yet clear.   

We thank Kevin Zahnle, Joe Harrington, David Charbonneau, and Sara Seager for comments.  We acknowledge support from an NRC postdoctral fellowship (JJF), NASA grants NAG2-6007 and NAG5-8919 (MSM), and NSF grant AST-0406963 (KL).  This work was supported in part by the US Department of Energy under contract W-7405-ENG-36.

\newpage

\newpage

\newpage
\begin{deluxetable}{cccccccc} 
\center
\tablecolumns{8}
\tablewidth{0pc}
\tablecaption{Model Flux Density Ratios in Spitzer Bands for \T~and \hd \label{table:param}}
\tablehead{
\colhead{Planet} & \colhead{Model}   & \colhead{$T_{\mathrm{eff}}$} & \colhead{3.6$\mu$m} &
\colhead{4.5$\mu$m} & \colhead{5.8$\mu$m}   & \colhead{8.0$\mu$m}    & \colhead{24$\mu$m}}
\startdata 
\T  & 4$\pi$ & 1134 & 0.46 & 0.81 & 1.11 & 1.48 & 2.93\\ 
\T  & 2$\pi$ & 1352 & 1.06 & 1.30 & 1.69 & 2.24 & 3.69\\ 
\T  & 4$\pi$,3.2$\times$solar & 1144 & 0.58 & 0.78 & 1.15 & 1.62 & 3.02\\ 
\T  & 4$\pi$,5$\times$solar & 1148 & 0.61 & 0.78 & 1.17 & 1.66 & 3.06\\ 
\T  & 4$\pi$,heating & 1165 & 0.56 & 0.90 & 1.24 & 1.71 & 3.19\\ 
\hd & 4$\pi$ & 1442 & 0.96 & 1.12 & 1.45 & 1.90 & 3.05\\
\hd & 4$\pi$,cloudy & 1448 & 1.02 & 1.22 & 1.56 & 2.01 & 3.16\\ 
\enddata
\tablecomments{All ratios have been multiplied by 1000.  Abundances are solar, unless noted. \hd~``cloudy'' includes enstatite and iron clouds.}
\end{deluxetable}

\end{document}